# Adequacy of the Dicke model in cavity QED: a counter-"no-go" statement

András Vukics and Peter Domokos

Institute for Solid State Physics and Optics,
Wigner Research Centre for Physics, Hungarian Academy of Sciences,
P.O. Box 49, H-1525 Budapest, Hungary
`vukics.andras@wigner.mta.hu`

7th June 2012

## Abstract

The long-standing debate whether the phase transition in the Dicke model can be realized with dipoles in electromagnetic fields is yet an unsettled one. The well-known statement often referred to as the "no-go theorem", asserts that the so-called A-square term, just in the vicinity of the critical point, becomes relevant enough to prevent the system from undergoing a phase transition. At variance with this common belief, in this paper we prove that the Dicke model does give a consistent description of the interaction of light field with the internal excitation of atoms, but in the dipole gauge of quantum electrodynamics. The phase transition cannot be excluded by principle and a spontaneous transverse-electric mean field may appear. We point out that the single-mode approximation is crucial: the proper treatment has to be based on cavity QED, wherefore we present a systematic derivation of the dipole gauge inside a perfect Fabry-Pérot cavity from first principles. Besides the impact on the debate around the Dicke phase transition, such a cleanup of the theoretical ground of cavity QED is important because currently there are many emerging experimental approaches to reach strong or even ultrastrong coupling between dipoles and photons, which demand a correct treatment of the Dicke model parameters.

PACS: 05.30.Rt,37.30.+i,42.50.Nn,42.50.Pq

# Contents

# 1 Introduction

The Dicke model describes an ensemble of two-level systems interacting with a radiation mode [1–3]:

$$H_{\text{Dicke}} = \omega_A S_z + \omega_C a^\dagger a + \frac{y}{\sqrt{N}} \left( a + a^\dagger \right) S_x, \qquad (1)$$

where the spin represents a collective degree of freedom of the two-level systems and $a, a^\dagger$ are the bosonic mode operators. It was Hepp and Lieb [4], and, simultaneously, Wang and Hioe [5], who showed that the Dicke model features a first-order thermal phase transition. On including the counter-rotating terms as stands in Eq. (1), the transition becomes of second order [6]. Beyond a critical coupling strength $y_c$, there is a critical temperature at which the system goes into a superradiant phase. This thermal phase transition has a zero-temperature counterpart at $y_c = \sqrt{\omega_C \omega_A}$, which has been further elaborated in a number of studies [7, 8], also in connection with entanglement properties [9–11].

In a parallel development, it was shown that as far as the spin represents the atomic internal electronic degree of freedom, the Dicke model is a fictitious one [12–16], because it is based on neglecting the $A^2$ term which stems from the kinetic part of the Hamiltonian. If this term is included, it becomes dominant enough in the vicinity of the critical point to wipe away the phase transition. Again, it is important to emphasize that this no-go statement pertains to the *derivation* of the Dicke model from the standard Coulomb-gauge Hamiltonian, but it does not concern any results derived from the Dicke model itself. From the perspective of this latter type of studies, the only question remains whether this model can be realized with any other physical system than electric dipoles in electromagnetic fields considered in the Coulomb gauge.

Indeed, the model has been applied in nuclear physics [17], while much more recently in circuit cavity QED [18] and cavity QED with motional excitations of a Bose-Einstein condensate [19]. These applications, given that their starting points are different from the one discussed here, are quite independent of the debate addressed in the present paper. In the latter case the phase transition was in fact experimentally demonstrated [20].

A significant contribution to the debate came recently from Keeling, who pointed out [21] that complementing the Dicke model with terms stemming from the $A^2$ term is still not consistent because a further term of equal importance must be taken into account: the instantaneous Coulomb interaction between the charges belonging to different atoms. His conclusion was that with this term included in the model Hamiltonian (which hence has the form of a complemented Dicke model), the criticality is restored. It was shown that transforming the model Hamiltonian into the other obvious choice of gauge, the electric-dipole gauge (cf. [22] chapter IV.), *one exactly recovers the original Dicke model* (1), with the modified interpretation that in this gauge, the boson mode $a$ corresponds not to a single mode of the transverse electric field, but that of the displacement field. It is the displacement field that undergoes spontaneous symmetry breaking which is shown to be due to atomic polarization only, the transverse electric field – a gauge-invariant quantity – remaining zero also in the electric-dipole gauge.

A serious limitation of the approach of Ref. [21], however, is that there is an arbitrary picking-out of a single traveling-wave mode, which pertains not only to the electromagnetic field but also a single spatial mode of the dipole-dipole interaction. The physical picture behind this is rather hard to see: because even if we assume that a cavity will select a single mode of the field, it will not reduce the dipole-dipole interaction to this same spatial mode. Rather, the restriction to a single spatial mode can be interpreted as replacing each dipole by a (complex) polarization density spread evenly over the whole space. This implies that in this approach, we in fact cannot distinguish between electric field and displacement, anywhere in space.

In this paper we propose to describe localized atoms interacting with the electromagnetic field confined in a resonator which has *true* discrete modes. To be specific, we choose the simplest possible geometry, the ideal, one-dimensional Fabry-Pérot resonator. We can thus get rid of the problem of selecting a single mode that arises in the infinite free-space case, which proved to be the source of the controversial consequences of usually neglected terms, such as the $A^2$. To describe the interaction of atoms with the radiation field, our starting point is the Coulomb-gauge minimal-coupling Hamiltonian, which we adapt to the specific resonator boundary conditions. Then we perform the canonical transformation into the electric-dipole gauge. We obtain a Hamiltonian of the Dicke type in Eq. (1) without any extra term. This leads us to conclude that the phase transition (i.e. the appearance of a spontaneous transverse-electric mean field) is not by any principle excluded. The critical point requires such a large dipole coupling which in fact corresponds to the ultrastrong coupling limit of cavity QED [23]. While it holds true that this may be difficult to realize with atoms, emerging new systems do seem to reach this regime [24, 25].

To the best of our knowledge, the canonical derivation of the electric-dipole gauge for a resonator geometry has not been systematically carried out before. Going beyond the special generic case, we can conjecture that the formalism of cavity QED can be founded on the multipolar-coupling scheme of QED. Since the two descriptions, minimal vs. multipolar coupling, are connected by canonical transformation, they are equivalent and must yield the same value for all physical observables. However, further approximations usually invoked to describe given experimental configurations, such as the ubiquitous two-level (or few-level) approximation, are not gauge invariant [26]. Hence the very definition of a (two-level) atom may depend on the choice of gauge. The possibility of using the electric-dipole gauge for cavity QED is essential since it treats the atom-atom interaction (if atoms are separated by more than a wavelength) as mediated exclusively by the transverse radiation field (the displacement) in conformity with the retardation principle. Less fundamental but important is that, in this gauge, the canonical momentum coincides with the kinetic momentum, making the description of mechanical effects of light on atoms within a cavity, like cavity cooling, direct [27]. Therefore, the forthcoming analysis concerns the foundations of cavity QED, in general, showing that it is compatible with the electric-dipole gauge.

The paper is organized as follows: After sketching the layout of the system in Section 2, in Section 3 we derive its minimal-coupling Hamiltonian. In Section 4 we show an outline of the derivation of our principal result, the transformation to electric-dipole gauge, which results in such a Hamiltonian of purely retarded form as is compatible with the Dicke model (Section 5). Many details of the calculation are presented in the Appendices.

## 2  Geometry, boundary conditions, and mode structure

We assume two parallel infinite planes of perfect conductors in the $x-y$ plane. Hence, the axis of our Fabry-Pérot cavity lies in the $z$ direction. The left mirror is situated at $z = 0$. We denote the length of the cavity by $L$. The layout is sketched in Fig. 1.

The boundary conditions are:

$$\hat{\mathbf{z}} \times \mathbf{E} = 0 = \hat{\mathbf{z}} \mathbf{B}, \quad \text{at the mirrors.} \tag{2}$$

We define the mode functions satisfying these boundary conditions [28]:

$$\boldsymbol{\psi}_{\mathbf{k}}^{\mathrm{E}}(\mathbf{r}) \equiv \hat{\mathbf{k}}_{\perp} \times \hat{\mathbf{z}} \sin(k_n z) e^{i \mathbf{k}_{\perp} \mathbf{r}_{\perp}}, \tag{3a}$$

$$\boldsymbol{\psi}_{\mathbf{k}}^{\mathrm{M}}(\mathbf{r}) \equiv \frac{1}{k} \left( k_{\perp} \cos(k_n z) \hat{\mathbf{z}} - i k_n \sin(k_n z) \hat{\mathbf{k}}_{\perp} \right) e^{i \mathbf{k}_{\perp} \mathbf{r}_{\perp}}, \tag{3b}$$

designated transverse electric (TE) and magnetic (TM) modes, respectively, with reference to the direction of the fields related to the $\mathbf{k}$ vector. We have introduced

$$\mathbf{r} \equiv z \hat{\mathbf{z}} + \mathbf{r}_{\perp}, \quad \mathbf{k} \equiv k_n \hat{\mathbf{z}} + \mathbf{k}_{\perp}, \quad \mathbf{r}_{\perp}, \mathbf{k}_{\perp} \perp \hat{\mathbf{z}}, \quad k_n \equiv \frac{n\pi}{L}. \tag{4}$$

Note that the polarizations are fixed by the direction of $\mathbf{k}_{\perp}$, while in the case of $\mathbf{k}_{\perp} = 0$ we are left with two sinusoidal modes of orthogonal polarization, which can be chosen arbitrarily in the transverse direction of the cavity. Importantly, there is no TE mode corresponding to $k_n = 0$, only TM mode; this is observed throughout when summation over $n$ enters the calculations, even if not expressly stated.

The dispersion relation of the modes reads:

$$\omega_k^2 \equiv c^2 \left( k_n^2 + k_{\perp}^2 \right) \equiv c^2 k^2. \tag{5}$$

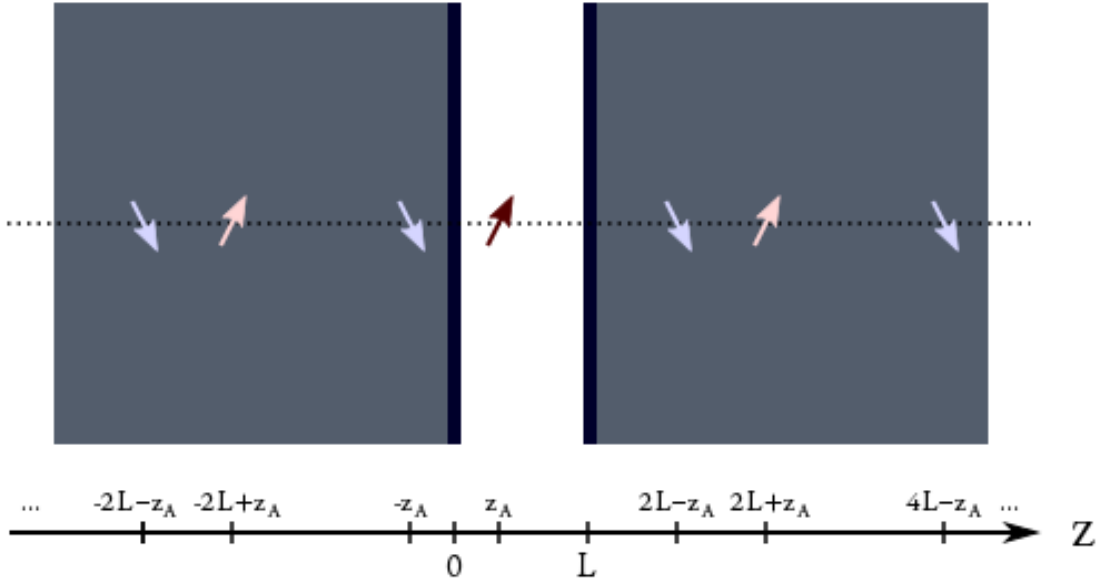

Figure 1: The geometry of our Fabry-Pérot cavity containing a single dipole. The image dipoles are indicated with light gray. For the corresponding distribution cf. Eq. (9).

In connection with the modes it is worth noting, moreover, that even in the case of all the dipoles distributed in line with the z axis and aligned transversely, for our purposes here it is necessary to take into account the *complete* mode structure of the field.

## 3   Dipole self-energy and dipole-dipole interaction in the Coulomb potential

Our starting point is the minimal-coupling (Coulomb-gauge) Hamiltonian of quantum electrodynamics *with point charges* (modeled after the one in [22] p. 173):

$$H_{\text{minimal-coupling}} = \sum_\alpha \frac{1}{2m_\alpha} [\mathbf{p}_\alpha - q_\alpha \mathbf{A}(\mathbf{r}_\alpha)]^2 + V_{\text{Coulomb}} + H_{\text{field}}, \tag{6a}$$

with

$$H_{\text{field}} \equiv \int d^3k \, \hbar\omega_k \left( a_{\mathbf{k}}^{E\dagger} a_{\mathbf{k}}^{E} + a_{\mathbf{k}}^{M\dagger} a_{\mathbf{k}}^{M} + 1 \right). \tag{6b}$$

The boundary conditions enter at two places:

1. In the Coulomb term because the longitudinal field must satisfy the boundary conditions.

2. In the mode structure of the field as each mode also must satisfy the boundary conditions individually.

Technically speaking, it is the purpose of this paper to show that these two appearances of the boundary condition conspire in such a way that in electric-dipole gauge they cancel each other, to yield the same *form* of Hamiltonian as in free space.

The modal decomposition of the vector potential reads:

$$\mathbf{A}(\mathbf{r}) = \mathbf{A}^E(\mathbf{r}) + \mathbf{A}^M(\mathbf{r}) = \int d^3k \sqrt{\frac{\hbar}{\varepsilon_0 \omega_k V}} \left( \boldsymbol{\psi}_{\mathbf{k}}^{E}(\mathbf{r}) a_{\mathbf{k}}^{E} + \boldsymbol{\psi}_{\mathbf{k}}^{M}(\mathbf{r}) a_{\mathbf{k}}^{M} \right) + \text{H.c.}. \tag{6c}$$

As indicated by Eq. (6b), the integral measure is dimensionless, and in the $z$ direction it is actually defined only as a summation:

$$\fint d^3 k\, f(\mathbf{k}) \equiv \frac{S}{(2\pi)^2} \sum_{n=0}^{\infty} \int d^2 k_\perp\, f(k_n \hat{\mathbf{z}} + \mathbf{k}_\perp), \tag{6d}$$

for any function $f$, where $S$ is the surface of the mirrors and the $(2\pi)^{-2}$ factor enters in a similar way as in [22] p. 31. Let us note once more that in connection with TE modes, the $n = 0$ term should be omitted from such sums.

In the presence of electric dipoles, the Coulomb term will be determined by the method of image charges. From this point on, we restrict ourselves to the case of neutral charge clusters (e. g. atoms) separated from each other by much larger distances than the cluster sizes. The dipole moment of cluster $A$ is the (quantum-mechanical) observable

$$\mathbf{d}_A \equiv \sum_{\alpha \in A} q_\alpha \mathbf{r}_\alpha. \tag{7}$$

The Coulomb term will be composed of three parts: (1) dipole self-interaction, which in our case appears in the minimal-coupling gauge as well, due to the presence of image dipoles; (2) dipole-dipole interaction; (3) the internal Coulomb energy of the clusters, wherein we discard the (infinite) Coulomb self-energy:

$$V_{\text{Coulomb}} \equiv V_{\text{Coulomb}}^{\text{dipole-self}} + V_{\text{Coulomb}}^{\text{dipole-dipole}} + \sum_A \sum_{\alpha \in A} \frac{q_\alpha}{8\pi\varepsilon_0} \sum_{\substack{\beta \in A \\ \beta \neq \alpha}} \frac{q_\beta}{|\mathbf{r}_\alpha - \mathbf{r}_\beta|}. \tag{8}$$

Putting a single dipole into the cavity, the field inside is the same as if it was created by the original dipole plus a series of image dipoles with alternating direction outside the cavity (cf. Fig. 1). Such a series can be described with the polarization density:

$$\mathbf{P}(z) = \sum_{n=-\infty}^{\infty} \left[ \delta\left(2nL + z_{\text{dip}} - z\right) \mathbf{1} + \delta\left(2nL - z_{\text{dip}} - z\right) \mathbf{R} \right] \mathbf{d}, \tag{9}$$

where $z_{\text{dip}}$ is the position of the real dipole, and the first term of the $n = 0$ term in the series corresponds to this. We have introduced the reflection operator

$$\mathbf{R} \equiv \begin{pmatrix} -1 & 0 & 0 \\ 0 & -1 & 0 \\ 0 & 0 & 1 \end{pmatrix}. \tag{10}$$

We obtain the following self-interaction term:

$$V_{\text{Coulomb}}^{\text{dipole-self}} = \frac{1}{8\pi\varepsilon_0 L^3} \sum_A \mathbf{d}_A \left[ \frac{\zeta(3)}{4} \begin{pmatrix} 1 & 0 & 0 \\ 0 & 1 & 0 \\ 0 & 0 & -2 \end{pmatrix} + \xi\left(\frac{2z}{L}, 0\right) \begin{pmatrix} -1 & 0 & 0 \\ 0 & -1 & 0 \\ 0 & 0 & -2 \end{pmatrix} \right] \mathbf{d}_A, \tag{11}$$

where $z$ is the dipole's distance from one of the mirrors. Besides Apéry's constant $\zeta(3)$ ($\zeta$ being the Euler-Riemann function) we also need to rely on the function

$$\xi(u, v) \equiv \sum_n \left((2n + u)^2 + v^2\right)^{-\frac{3}{2}} \tag{12}$$

throughout, which in the case of $v = 0$ remains finite if $0 < u < 2$. We note in passing that $V_{\text{Coulomb}}^{\text{dipole-self}}$ breaks the spherical symmetry of the atomic Hamiltonian, and hence the Thomas-Reiche-Kuhn sum rule, which played an essential role in the no-go argumentation of [12], is not applicable in this case. (We are not elaborating on this point here, however, since our concern is not the Dicke model's validity *in the minimal-coupling gauge*.)

The presence of image dipoles rises once more the problem of retardation. The additional field inside the cavity attributed to image dipoles is in reality produced by a surface-charge density on the mirrors, which arises because of the jump of the electric field's normal component at the boundary. On the real dipole's motion within the cavity, the surface charges undergo an *instantaneous* redistribution. Again, it is expected that this instantaneous effect gets canceled by the radiation field, which also satisfies a non-trivial boundary condition at the mirrors.

A second dipole placed into the cavity will interact not only with the other real dipole, but also the latter's full series of images. This yields an interaction term depending not only on the position difference of the two dipoles, but also on the position of their center of mass relative to one of the mirrors:

$$V_{\text{Coulomb}}^{\text{dipole-dipole}} = \frac{1}{8\pi\varepsilon_0} \sum_{\substack{A,B \\ A \neq B}} \mathbf{d}_A \left[ \mathfrak{C}^{(+)}(\delta\mathbf{r}_{A-B}) + \mathfrak{C}^{(-)}\left((z_A + z_B)\hat{\mathbf{z}} + \delta\mathbf{r}_{\perp A-B}\right) \right] \mathbf{d}_B. \tag{13}$$

The arising matrix is found to read

$$\mathfrak{C}^{(+)}(\boldsymbol{\rho}) = \sum_n \frac{1}{\rho_n^3} \begin{pmatrix} 1 - 3\frac{\rho_\perp^2}{\rho_n^2} & 0 & -3\frac{\rho_\perp(2Ln+\rho_z)}{\rho_n^2} \\ 0 & 1 & 0 \\ -3\frac{\rho_\perp(2Ln+\rho_z)}{\rho_n^2} & 0 & 1 - 3\frac{(2Ln+\rho_z)^2}{\rho_n^2} \end{pmatrix}, \tag{14a}$$

where

$$\boldsymbol{\rho} \equiv \begin{pmatrix} \rho_\perp \\ 0 \\ \rho_z \end{pmatrix}, \quad \rho_n^2 \equiv (2Ln + \rho_z)^2 + \rho_\perp^2. \tag{14b}$$

In the more general case of an arbitrary $\boldsymbol{\rho}$ direction (as will be the case with arbitrarily placed dipoles), the matrix must be suitably rotated, whereupon the zero elements become nonzero as well. The complementary matrix is found to be related simply as

$$\mathfrak{C}^{(-)} = \mathfrak{C}^{(+)}\mathbf{R}. \tag{14c}$$

The full minimal-coupling Hamiltonian then reads:

$$H_{\substack{\text{minimal-coupling} \\ \text{with electric dipoles}}} = \sum_A \sum_{\alpha \in A} \left\{ \frac{[\mathbf{p}_\alpha - q_\alpha \mathbf{A}(\mathbf{r}_A)]^2}{2m_\alpha} + \frac{q_\alpha}{8\pi\varepsilon_0} \sum_{\substack{\beta \in A \\ \beta \neq \alpha}} \frac{q_\beta}{|\mathbf{r}_\alpha - \mathbf{r}_\beta|} \right\} + V_{\text{Coulomb}}^{\text{dipole-self}} + V_{\text{Coulomb}}^{\text{dipole-dipole}} + H_{\text{field}}, \tag{15}$$

where in the first line we separated the single-dipole Hamiltonian, in whose first term we have already used the dipole (long-wavelength) approximation $\mathbf{A}(\mathbf{r}_\alpha) \to \mathbf{A}(\mathbf{r}_A)$, keeping only the first term in the multipolar series.

## 4 Canonical transformation to electric-dipole gauge

It is the Hamiltonian (15) that we transform into the electric-dipole gauge, by applying to it the usual transformation

$$U = \exp\left\{ -\frac{i}{\hbar} \sum_A \mathbf{d}_A \mathbf{A}(\mathbf{r}_A) \right\}$$

$$= \exp\left\{ -\frac{i}{\hbar} \sum_A \sum_{\alpha \in A} \mathbf{r}_\alpha \left[ q_\alpha \mathbf{A}(\mathbf{r}_A) \right] \right\} = \exp\left\{ \int d^3k \left( \lambda_{\mathbf{k}}^{\text{E}} \right)^* a_{\mathbf{k}}^{\text{E}} + \left( \lambda_{\mathbf{k}}^{\text{M}} \right)^* a_{\mathbf{k}}^{\text{M}} - \text{H.c.} \right\}, \tag{16a}$$

which touches only the kinetic and the $H_{\text{field}}$ part of the Hamiltonian (15). The transformation makes sense in finite space since it does not modify the position variable. From the forms in the second line it is clear that $U$ amounts, on one hand, to shifting the canonical momentum $\mathbf{p}_\alpha$ in the Hamiltonian, so that the canonical momentum coincides with the kinetic momentum in the new picture, and, on the other hand, to shifting the annihilation operators by

$$\lambda_{\mathbf{k}}^{\mathrm{E,M}} = \frac{i}{\sqrt{\hbar \varepsilon_0 \omega_k V}} \sum_A \mathbf{d}_A \left( \psi_{\mathbf{k}}^{\mathrm{E,M}}(\mathbf{r}_A) \right)^* . \tag{16b}$$

Out of the $H_{\mathrm{field}}$ part of the Hamiltonian (15), two kinds of terms appear. Firstly, terms linear in the $\lambda_{\mathbf{k}}^{\mathrm{E,M}}$s, which will describe the atom-field interaction in this gauge. Secondly, we have the quadratic terms

$$\fint d^3 k\, \hbar \omega_k \left[ \left( \lambda_{\mathbf{k}}^{\mathrm{E}} \right)^* \lambda_{\mathbf{k}}^{\mathrm{E}} + \left( \lambda_{\mathbf{k}}^{\mathrm{M}} \right)^* \lambda_{\mathbf{k}}^{\mathrm{M}} \right]. \tag{17}$$

This is composed of terms containing only one dipole moment and terms containing two dipole moments.

## 4.1 The two-dipole terms

Let us first consider the terms containing two dipole moments. These can be written with the help of a matrix again as:

$$H_{\mathrm{quadratic}}^{\mathrm{dipole\text{-}dipole}} = \frac{1}{16\pi^2 \varepsilon_0 L} \sum_{\substack{A,B \\ A \neq B}} \mathbf{d}_A \left[ \mathfrak{D}^{(+)}(\delta \mathbf{r}_{A-B}) + \mathfrak{D}^{(-)} \left( (z_A + z_B) \hat{\mathbf{z}} + \delta \mathbf{r}_{\perp A-B} \right) \right] \mathbf{d}_B. \tag{18}$$

For brevity, the arising matrix we list already after the azimuthal integration, which invokes the first three of the cylindrical Bessel functions $J_n$:

$$\mathfrak{D}^{(+)}(\boldsymbol{\rho}) = \pi \sum_n e^{i k_n \rho_z} \int_0^\infty dk_\perp \frac{k_\perp}{k^2}$$
$$\cdot \begin{pmatrix} (2k_n^2 + k_\perp^2) J_0(k_\perp \rho_\perp) + k_\perp^2 J_2(k_\perp \rho_\perp) & 0 & -2i k_n k_\perp J_1(k_\perp \rho_\perp) \\ 0 & (2k_n^2 + k_\perp^2) J_0(k_\perp \rho_\perp) - k_\perp^2 J_2(k_\perp \rho_\perp) & 0 \\ -2i k_n k_\perp J_1(k_\perp \rho_\perp) & 0 & 2k_\perp^2 J_0(k_\perp \rho_\perp) \end{pmatrix}. \tag{19a}$$

We have used that due to symmetry $\int_0^{2\pi} d\phi\, \sin(\phi)\, e^{i x \cos(\phi)} = 0 = \int_0^{2\pi} d\phi\, \sin(\phi) \cos(\phi)\, e^{i x \cos(\phi)}$. The matrix $\mathfrak{D}^{(+)}$ much resembles the transverse delta function (cf. Eq. (26)). In contrast to the free-space case (cf. [22] pp. 36–44), however, in our case $\mathfrak{D}^{(+)}(\boldsymbol{\rho})$ remains finite for nonzero $\boldsymbol{\rho}$, without introducing any ultraviolet cutoff. Concerning the direction of $\boldsymbol{\rho}$ the same note applies as in the paragraph after Eq. (14). The complementary matrix is here again found to be related simply as

$$\mathfrak{D}^{(-)} = \mathfrak{D}^{(+)} \mathbf{R}. \tag{19b}$$

It is important to note that in the case of axial alignment

$$\mathrm{D}_{xz}^{(+/-)}(\rho_z, \rho_\perp = 0) = 0 = \mathfrak{D}_{xz}^{(+/-)}(\rho_z, \rho_\perp = 0) \quad \forall \rho_z, \tag{20}$$

the matrices becoming invariant under rotations around the cavity axis.

Recall that in the free-space case (where only a term similar to the $\mathfrak{D}^{(+)}$ term appears), the (18) type of term cancels the instantaneous Coulomb interaction between two dipoles, yielding a Hamiltonian of purely retarded form. Here we aimed at proving that this is extended to the whole of $V_{\mathrm{Coulomb}}^{\mathrm{dipole\text{-}dipole}}$. Indeed, the structure of (13) is very similar to (18). The identity that we wish to prove reads

$$\mathfrak{C}^{(+)}(\boldsymbol{\rho}) \stackrel{?}{=} -\frac{1}{2\pi L} \mathfrak{D}^{(+)}(\boldsymbol{\rho}) \quad \forall \boldsymbol{\rho}. \tag{21}$$

Some details of the calculation are given in Appendix A. We have found (cf. Appendix B) that this matrix identity can be further distilled to the stronger identity

$$\int_0^\infty dx \frac{x \cosh(x[u-1])}{\sinh(x)} J_1(xv) \stackrel{?}{=} v\, \xi(u,v) \quad \forall\, 0 < u < 2,\, v, \tag{22}$$

for which we offer a straightforward proof in Appendix C.

We have proven thereby that in a Fabry-Pérot resonator, too, the electric-dipole Hamiltonian is of a fully retarded form, meaning in this case that the transformation (16a) cancels not only the instantaneous dipole-dipole interaction terms of the Hamiltonian (15), but the dipole's interaction with another dipole's *whole series of images* as well.

## 4.2 The single-dipole terms

The quadratic term containing only terms of a single dipole moment reads

$$H^{\text{dipole-self}}_{\text{quadratic}} = \frac{1}{16\pi^2 \varepsilon_0 L} \sum_A \mathbf{d}_A \left[ \mathfrak{D}^{(+)}(0) + \mathfrak{D}^{(-)}(2z_A) \right] \mathbf{d}_A. \tag{23}$$

Relying on the identity (22) (or, more precisely, the weaker identity (29)), the second term is easily proven to cancel the second term of $V^{\text{dipole-self}}_{\text{Coulomb}}$ defined in Eq. (11), leaving us with a position-independent dipole-self energy.

Characterization of the first term requires much greater care because the $\mathfrak{D}$ matrices are ill-defined at the origin. In Appendix D we offer a proof that $\mathfrak{D}^{(+)}$ is isotropic for large $L$, when the summation can be replaced by integration. Our conclusion then is that in dipole gauge, the dipole-self energy $E^{\text{dipole-self}}$ has an isotropic but infinite part (just like we saw in the free-space case, cf. [22] p. 312), supplemented by a small anisotropic term vanishing as $L^{-3}$ with increasing cavity length.

## 5 Conclusion: the Dicke model from the electric-dipole gauge

The total electric-dipole Hamiltonian is then the Hamiltonian (15) with the simplified kinetic term, plus $H^{\text{dipole-dipole}}_{\text{quadratic}} + H^{\text{dipole-self}}_{\text{quadratic}}$, supplemented by the term linear in the $\lambda^{\text{E,M}}_{\mathbf{k}}$s describing the dipole coupling:

$$H_{\text{electric-dipole}} = \sum_A \left\{ \sum_{\alpha \in A} \left[ \frac{\mathbf{p}_\alpha^2}{2m_\alpha} + \frac{q_\alpha}{8\pi\varepsilon_0} \sum_{\substack{\beta \in A \\ \beta \neq \alpha}} \frac{q_\beta}{|\mathbf{r}_\alpha - \mathbf{r}_\beta|} \right] + E_A^{\text{dipole-self}} - \mathbf{d}_A \frac{\mathbf{D}(\mathbf{r}_A)}{\varepsilon_0} \right\} + H_{\text{field}}, \tag{24a}$$

where the annihilation operators are coefficients not of the electric, but the (transverse) displacement field:

$$\frac{\mathbf{D}_\perp(\mathbf{r})}{\varepsilon_0} = i \int d^3k \sqrt{\frac{\hbar \omega_k}{\varepsilon_0 V}} \left( \boldsymbol{\psi}^{\text{E}}_{\mathbf{k}}(\mathbf{r})\, a^{\text{E}}_{\mathbf{k}} + \boldsymbol{\psi}^{\text{M}}_{\mathbf{k}}(\mathbf{r})\, a^{\text{M}}_{\mathbf{k}} \right) - \text{H.c.}. \tag{24b}$$

From this Hamiltonian such approximations lead to the Dicke model (1) as are not questioned by any no-go statement. These are (note that the dipole approximation is already inherent in our whole treatment)

- the two-level approximation for the atoms, replacing the first two terms of the Hamiltonian (24a) with the first term of the Dicke Hamiltonian (1), with the anisotropy of $E^{\text{dipole-self}}$ perturbing the value of $\omega_A$ in comparison to the free-space case; and

- the single-mode approximation for the field, justified by the presence of the cavity, replacing the last two terms of the Hamiltonian (24a) with the last two terms of (1).

There remains to clarify the connection between the transverse displacement field $\mathbf{D}_\perp$ and the transverse electric field $\mathbf{E}_\perp$, to show what the effect of a spontaneous mean field in the selected mode of the former beyond the transition point on the latter is. Since every dipole is surrounded by a cloud of $\mathbf{P}_\perp(\mathbf{r})$ and $\mathbf{P}_\parallel(\mathbf{r})$ fields (of opposite signs), both vanishing inverse cubically with increasing distance from the dipole, a mean field in a mode of $\mathbf{D}_\perp$ (whose spatial distribution is determined by the mode function alone) might be carried by $\mathbf{P}_\perp$ *only around the dipoles*, while far from the dipoles its carrier must be $\mathbf{E}_\perp$. This means that the phase transition in the Dicke model that we obtained in *electric-dipole gauge* must result in a spontaneous mean field in the *gauge-invariant* observable $\mathbf{E}_\perp$.

From our results it is by no means far-fetched to conjecture that in cases of more generic boundary conditions, as in actual cavity experiments, similar cancellation of instantaneous dipole-dipole and dipole-image interaction terms occurs when transforming to electric-dipole gauge. This means that the fully retarded Hamiltonian (24a) is the general form for cavity QED in electric-dipole gauge, the boundary conditions entering only through the modal decomposition of the displacement field.

# A  The mode structure compared to the free-space case

To better understand the connection with the free-space case, one may rewrite the mode functions (3) as follows. On writing sin and cos with exponentials, and merging an $i$ phase factor into the $a^E$ operators, we obtain the form

$$\mathbf{A}(\mathbf{r}) = \frac{1}{2} \oint d^3k \sqrt{\frac{\hbar}{\varepsilon_0 \omega_k V}} \left( \hat{\mathbf{k}}_\perp \times \hat{\mathbf{z}}\, a_\mathbf{k}^E + \boldsymbol{\epsilon}_\mathbf{k} a_\mathbf{k}^M \right) e^{i\mathbf{k}\mathbf{r}} + \text{H.c.}, \tag{25a}$$

where

$$\oint d^3k\, f(\mathbf{k}) \equiv \frac{S}{(2\pi)^2} \sum_{n=-\infty}^{\infty} \int d^2k_\perp f(k_n \hat{\mathbf{z}} + \mathbf{k}_\perp), \quad \boldsymbol{\epsilon}_\mathbf{k} = \frac{1}{k}\left( k_\perp \hat{\mathbf{z}} - k_n \hat{\mathbf{k}}_\perp \right). \tag{25b}$$

$\boldsymbol{\epsilon}_\mathbf{k}$ is orthogonal to both $\mathbf{k}$ and $\hat{\mathbf{k}}_\perp \times \hat{\mathbf{z}}$. Differences from the free-space case:

1. $k_z$ integral is replaced by summation;

2. annihilation operators corresponding to opposite-sign $k_z$s are related $a_\mathbf{k} = -a_{-\mathbf{Rk}}$, which explains the additional 1/2 factor in the normalization of (25a);

3. the polarization of the modes is not arbitrary in the direction orthogonal to $\mathbf{k}$, but is fixed by the cavity-axis direction.

Using this form of the mode functions, the $\mathfrak{D}$ matrices can be written in very concise forms (omitting factors of $S$ and $\pi$):

$$\mathfrak{D}^{(+)}(\boldsymbol{\rho}) \propto \oint d^3k \left[ \boldsymbol{\epsilon}_\mathbf{k} \circ \boldsymbol{\epsilon}_\mathbf{k} + (\hat{\mathbf{k}}_\perp \times \hat{\mathbf{z}}) \circ (\hat{\mathbf{k}}_\perp \times \hat{\mathbf{z}}) \right] e^{i\mathbf{k}\boldsymbol{\rho}} = \oint d^3k \left( 1 - \hat{\mathbf{k}} \circ \hat{\mathbf{k}} \right) e^{i\mathbf{k}\boldsymbol{\rho}}, \tag{26a}$$

$$\mathfrak{D}^{(-)}(\boldsymbol{\rho}) \propto \oint d^3k \left[ \boldsymbol{\epsilon}_\mathbf{k} \circ \boldsymbol{\epsilon}_{-\mathbf{Rk}} - (\hat{\mathbf{k}}_\perp \times \hat{\mathbf{z}}) \circ (\hat{\mathbf{k}}_\perp \times \hat{\mathbf{z}}) \right] e^{i\mathbf{k}\boldsymbol{\rho}} = \mathfrak{D}^{(+)}(\boldsymbol{\rho})\, \mathbf{R}. \tag{26b}$$

Note that the $n = 0$ contribution stemming from the TE modes should be subtracted from these expressions. This would amount to subtracting a term $\frac{S}{(2\pi)^2} \int d^2k_\perp (\hat{\mathbf{k}}_\perp \times \hat{\mathbf{z}}) \circ (\hat{\mathbf{k}}_\perp \times \hat{\mathbf{z}})\, e^{i\mathbf{k}_\perp \boldsymbol{\rho}_\perp}$ from $\mathfrak{D}^{(+)}$ and adding the same term to $\mathfrak{D}^{(-)}$. Hence, these corrections cancel each other in expressions containing the sum of these two matrices taken at the same $\boldsymbol{\rho}_\perp$ point, as was always the case in this work.

From the forms (26) it is already straightforward to obtain the forms of Eq. (19). It is also apparent that $\mathfrak{D}^{(+)}$ is almost the transverse delta function (cf. [22] p. 38), the only difference being that for the $z$ component of the wave vector we have summation instead of integral.

# B  Proof of (21) following from (22)

In the expression (19) the summation can be performed with a technique similar to the one presented in [29] as

$$\sum_n e^{i\alpha n} \frac{n^m}{n^2 + \beta^2} = \oint_C dz\, e^{i\alpha z} \frac{z^m}{z^2 + \beta^2} \frac{1}{e^{2\pi i z} - 1} = \oint_S dz\, "$$

$$= 2\pi i \sum_{z=\pm i\beta} \operatorname*{Res}_z(") = \frac{\pi i^m \beta^{m-1}}{\sinh(\pi\beta)} \begin{cases} \cosh(\alpha\beta - \pi\beta) & \text{for even } m \\ \sinh(\alpha\beta - \pi\beta) & \text{for odd } m \end{cases} \tag{27}$$

For the corresponding contours consult Fig. 2. The sum is $2\pi$-periodic in $\alpha$, whence we can assume $0 \le \alpha < 2\pi$ (in our case $0 \le \rho_z < 2L$ is always true anyway). For nonzero $\alpha$, the integrand always vanishes for $|z| \to \infty$. In the important special case of $\alpha = 0$ (that is, $\rho_z = 0$), the sum

- yields a finite value for $m = 0$.

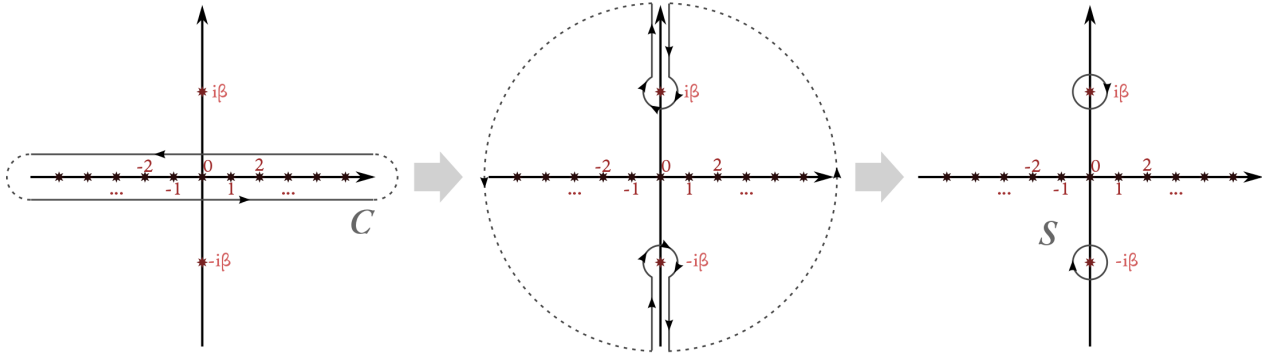

Figure 2: Distortion of the contour $C$ into $S$ for (27), with the integrand's poles indicated by stars. For the second step to be valid, it is required that the integrand vanish on the dotted section of the middle contour. For $0 < \alpha < 2\pi$, this is fulfilled for all $m$, while in the important special case of $\alpha = 0$, for $m < 2$.

- is zero for $m = 1$.
- diverges for $m \geq 2$.

Substitution yields:

$$\mathfrak{D}^{(+)}(\rho) = \frac{\pi}{L^2} \int_0^\infty dx \frac{x^2}{\sinh(x)} \begin{pmatrix} (-J_0(xv)+J_2(xv))\cosh(x(u-1)) & 0 & J_1(xv)\sinh(x(u-1)) \\ 0 & (-J_0(xv)-J_2(xv))\cosh(x(u-1)) & 0 \\ J_1(xv)\sinh(x(u-1)) & 0 & 2J_0(xv)\cosh(x(u-1)) \end{pmatrix}, \tag{28a}$$

where we have introduced

$$u \equiv \frac{\rho_z}{L}, \quad v \equiv \frac{\rho_\perp}{L}. \tag{28b}$$

Straightforward algebraic and analytic considerations yield that the four identities which stem from (21) for the matrices' four nonzero elements, and which we need to prove, are equivalent to the following three identities:

$$\int_0^\infty dx \frac{x^2 \cosh(x(u-1))}{\sinh(x)} [J_0(xv) \pm J_2(xv)] \stackrel{?}{=} [2 + (1 \mp 1) v\, \partial_v]\, \xi(u,v) \quad \text{and} \tag{29}$$

$$\int_0^\infty dx \frac{x^2 \sinh(x(u-1))}{\sinh(x)} J_1(xv) \stackrel{?}{=} v\, \partial_u \xi(u,v) \quad \forall\, 0 < u < 2, v. \tag{30}$$

Eq. (30) follows via a $u$-derivation from the stronger identity (22). Moreover, relying on the recurrence formulae (cf. [30] p. 16)

$$2\alpha J_\alpha(x) = x\left(J_{\alpha-1}(x) + J_{\alpha+1}(x)\right), \tag{31a}$$

$$2\frac{dJ_\alpha}{dx} = J_{\alpha-1}(x) - J_{\alpha+1}(x), \tag{31b}$$

Eqs. (29) also follow. Therefore, it is the single Eq. (22) that we need to prove, which is stronger than the desired identity Eq. (21).

## C  Proof of the identity (22)

We start from the expression of the reciprocal distance of two spatial points **r** and **r**′ in cylindrical coordinates (cf. [30] p. 102):

$$\frac{1}{|\mathbf{r}-\mathbf{r}'|} = \int_0^\infty d\lambda\, e^{-\lambda|z-z'|} J_0(\lambda\Delta r_\perp), \quad \Delta r_\perp = |\mathbf{r}_\perp - \mathbf{r}'_\perp| = \sqrt{r_\perp^2 + r_\perp'^2 - r_\perp r_\perp' \cos(\phi-\phi')}, \quad (32)$$

which we rewrite in dimensionless form

$$(u^2+v^2)^{-\frac{1}{2}} = \int_0^\infty d\varkappa\, e^{-\varkappa|u|} J_0(\varkappa v), \quad (33)$$

and differentiate through $v$ (otherwise in the following we cannot swap the integral and the summation), using that $J_0' = -J_1$:

$$v(u^2+v^2)^{-\frac{3}{2}} = \int_0^\infty d\varkappa\,\varkappa\, e^{-\varkappa|u|} J_1(\varkappa v). \quad (34)$$

We substitute $u \to 2n+u$ and sum over $n$ to recover the $\xi$ function (12) on the LHS:

$$v\,\xi(u,v) = \int_0^\infty d\varkappa\,\varkappa \left(\sum_n e^{-\varkappa|2n+u|}\right) J_1(\varkappa v)$$

$$= \int_0^\infty d\varkappa\,\varkappa \left(e^{-\varkappa u}\sum_{n=0}^\infty e^{-2\varkappa n} + e^{\varkappa u}\sum_{n=-\infty}^{-1} e^{2\varkappa n}\right) J_1(\varkappa v) = \int_0^\infty d\varkappa\,\varkappa \left(\frac{e^{-\varkappa u} + e^{\varkappa(u-2)}}{1 - e^{-2\varkappa}}\right) J_1(\varkappa v)$$

$$= \int_0^\infty d\varkappa\,\frac{\varkappa \cosh(\varkappa[u-1])}{\sinh(\varkappa)} J_1(\varkappa v) \quad \blacksquare \quad (35)$$

It is interesting to note that this expression is actually stronger than

$$\sum_n \left\{\left[(2n+u)^2+v^2\right]^{-\frac{1}{2}} - \left[(2n+u')^2+v^2\right]^{-\frac{1}{2}}\right\}$$

$$= \int_0^\infty d\varkappa\,\frac{\cosh(\varkappa[u-1]) - \cosh(\varkappa[u'-1])}{\sinh(\varkappa)} J_0(\varkappa v), \quad (36)$$

that can be obtained from writing Green's function of the Poisson equation for a space bounded by two parallel planes in cylindrical (cf. [30] p. 103) and Descartes coordinates. In this case, the terms of the subtractions on the two sides cannot be identified, for the simple reason that the integral of one term alone in the RHS does not converge.

## D  Characterizing $\mathfrak{D}^{(+)}(0)$

Let us now turn to the case of $\rho = 0$ required by Eq. (23), where we cannot rely on the form (28) because here the convergence of the sum (27) is problematic. Instead, we have to fall back to the form (19). We characterize the anisotropy of the matrix via the following quantity:

$$\Delta \equiv \mathfrak{D}_{xx}^{(+)}(0) - \mathfrak{D}_{zz}^{(+)}(0) = \pi \sum_n \int_0^\infty dk_\perp\,\frac{k_\perp}{k^2}(2k_n^2 - k_\perp^2)$$

$$= \frac{\pi^3}{L^2} \sum_n \int_0^\infty dx\,\frac{x}{x^2+n^2}(2n^2 - x^2). \quad (37)$$

To be able to tackle this expression, we should introduce a cutoff on the wave vector's modulus, otherwise either the sum (for the $k_n^2$ term) or the integral (for the $k_\perp^2$ term) does not converge. In the limit of large $L$, however, it is straightforward to prove that the anisotropy vanishes, independently of any cutoff. This is brought about by replacing the summation with an integral and using polar coordinates:

$$\Delta_{L\to\infty} \propto \int_{-\infty}^{\infty} dy \int_{0}^{\infty} dx \, \frac{x}{x^2+y^2}\left(2y^2 - x^2\right) = \left(\int_{0}^{\infty} dr \, r^2\right) \int_{0}^{\pi} d\phi \, \sin(\phi)\left(3\cos^2(\phi) - 1\right)$$

$$= \left(\int_{0}^{\infty} dr \, r^2\right) \int_{-1}^{1} du \left(3u^2 - 1\right) = \left(\int_{0}^{\infty} dr \, r^2\right) \left[u^3 - u\right]_{-1}^{1} = 0. \quad (38)$$